\documentclass[preprint]{imsart}

\usepackage{amsthm,amsmath,amssymb,natbib}
\usepackage{graphicx}
\RequirePackage[colorlinks,citecolor=blue,urlcolor=blue]{hyperref}

\startlocaldefs
\makeatletter
\renewcommand*\env@matrix[1][*\c@MaxMatrixCols c]{%
  \hskip -\arraycolsep
  \let\@ifnextchar\new@ifnextchar
  \array{#1}}
\makeatother

\newtheorem{lem}{Lemma}[section]
\newtheorem{prop}{Proposition}[section]
\newcommand{\tr}{\text{tr}}
\newcommand{\te}{\text{te}}
\newcommand{\cv}{\text{cv}}
\DeclareMathOperator*{\argmin}{arg\,min}
\endlocaldefs

\begin{document}

\begin{frontmatter}

\title{Selective inference after cross-validation}
\runtitle{Inference after cross-validation}

\author{\fnms{Joshua} \snm{Loftus}\corref{}\ead[label=e1]{joftius@stanford.edu}}
\address{\printead{e1}}
\affiliation{Stanford University}
\address{Department of Statistics\\Sequoia Hall\\390 Serra Mall\\Stanford, CA}

\runauthor{J. R. Loftus}

\begin{abstract}
This paper describes a method for performing inference on models chosen by cross-validation. When the test error being minimized in cross-validation is a residual sum of squares it can be written as a quadratic form. This allows us to apply the inference framework in Loftus et al. (2015) for models determined by quadratic constraints to the model that minimizes CV test error. Our only requirement on the model training procedure is that its selection events are regions satisfying linear or quadratic constraints. This includes both Lasso and forward stepwise, which serve as our main examples throughout. We do not require knowledge of the error variance $\sigma^2$. The procedures described here are computationally intensive methods of selecting models adaptively and performing inference for the selected model. Implementations are available in an R package.
\end{abstract}



\end{frontmatter}

\section{Introduction}
We consider a modeling scenario with outcome or response variable $y
\sim N(\mu, \sigma^2I)$ and a matrix $X$ of predictor variables. We
hope to model the response as $y = X\beta + \epsilon$ for some unknown
$\beta$, and imagine that $\beta$ is sparse in the sense that few of
its entries are nonzero. In this setting various model selection
procedures exist that select a subset of predictors $X_A$ with the
hope that $X_A\beta_A$ is a good approximation of $\mu$. For example,
forward stepwise begins with an empty model and then sequentially adds
the most predictive variable at each step. After $k$ steps the forward
stepwise model includes $k$ predictors. The Lasso estimator is defined
as 
\begin{equation}
  \label{eq:lasso}
  \hat \beta_\lambda := \argmin_\beta \| y - X\beta \|_2^2 + \lambda \| \beta \|_1
\end{equation}
and has the property that its solutions are increasingly sparse as
$\lambda$ increases. For large enough $\lambda$, $\beta_\lambda = 0$,
and if $\lambda$ decreases continuously then predictors will enter the
Lasso model one at a time in a similar fashion to forward stepwise. In
this way, both forward stepwise and Lasso algorithms can output models
in a sequence of increasing model complexity, i.e. number of estimated
parameters. 

For such model selection procedures one of the most important
questions in their practical application is {\em how complex of a
  model should we use?} How many steps should we allow forward
stepwise to take? What value of $\lambda$ will yield a reasonable
Lasso model? There are many answers to these questions motivated by
theory, but in practice cross-validation is perhaps the most widely
used. In $K$-fold cross-validation, the data are partitioned into $K$
independent subsets $D_1, \ldots, D_K$.. For each $k = 1,\ldots,K$,
subset $D_k$ is held out while the rest of the data are used to train
a sequence of predictive models $\hat f^k_\lambda$. The subscript
$\lambda$ indexes model complexity, and might take continuous values
for the Lasso, or integer values for forward stepwise. The predicted
value $\hat y^k_\lambda = \hat f^k_\lambda(x^k)$, and with a
predictive loss function $\ell(y, \hat y)$ the cross-validation
estimate of prediction error is given by 
\begin{equation}
  \label{eq:cvloss}
 \text{cv-Error}_{\ell}(\lambda) :=  \frac{1}{K} \sum_{k=1}^K
 \frac{1}{|D_k|} \ell (y^k, \hat y^k_\lambda) 
\end{equation}
Throughout this paper we use squared-error loss $\ell (y, \hat y)
= \|y - \hat y\|_2^2$ and suppress the $\ell$ notation unless
otherwise specified. Finally, to choose $\lambda$ it is reasonable to
pick the minimizer $\hat \lambda$ of \eqref{eq:cvloss} with the hope
that a model with complexity $\hat \lambda$ will have low prediction
error. To simplify notation we also assume $K$ divides $n$ and all
folds have equal size $|D_k| = n/K$, and consider the equivalent
problem of minimizing
\begin{equation}
  \label{eq:cvlosssimplified}
  \text{cv-Error}(\lambda) = \sum_{k=1}^K \| y^k - \hat y_\lambda^k \|_2^2.
\end{equation}

Much recent work on selective inference shows how to perform inference
for Lasso and forward stepwise when the model complexity $\lambda$ has
been specified independently of the data \cite{lee2013exact,
  taylor2013tests, tibshirani2014exact}. Principled choices of
$\lambda$ can be made if $\sigma^2$ is known. When $\sigma^2$ is
unknown, \cite{tian2015sqrtlasso} use the square-root Lasso of
\cite{sqrtlasso}, and \cite{gross2015internal, loftus2015groups} show
how to do selective $F$ tests. Cross-validation remains one of the
most popular methods in practice, and no previous work has shown how
to conduct inference conditional on models selected this way. We do
this now, leveraging the quadratic model selection framework described
in \cite{loftus2015groups}. 

\subsection{Selective hypothesis tests}

\cite{loftus2015groups} describe a framework for significance tests
adjusted for model selection when the choice of selected model is
determined by quadratic constraints. That is, denoting the selected
model as $M(y)$, we have for each possible model $m$
\begin{equation}
\begin{aligned}
  \label{eq:quaddecomp}
  M(y) = m \iff y \in \bigcap_{j \in J_m} \{ y : y^TQ_{m,j}y + a_{m,j}^Ty + b_{m,j} \geq 0 \}\\
\end{aligned}
\end{equation}
for some finite index set $J_m$, and $Q_{m,j}, a_{m,j}, b_{m,j}$
depend on $y$ only through $m$. This definition is general enough to
include the Lasso, marginal screening, forward stepwise, and forward
stepwise with groups of variables. In the present work we show that
cross-validation combined with any of the above also fit within this
framework.

Given a model $m$ there is an active set $A_m$ of variables included
in the model. For each $j \in A_m$ we wish to conduct a significance
test. Assuming
\begin{equation}
  \label{eq:normalmodel}
  y \sim N(\mu, \sigma^2I)
\end{equation}
we give a test for each of the parameters $\beta_{A_m,j}$ with
$j \in A_m$ where $\beta_{A_m}$ is the population least squares
parameter vector $P_{A_m}\mu$, the projection of $\mu$ onto the
column space of $X_{A_m}$. If the linear model is correctly specified
and $A_m$ contains the true active set, then the null hypotheses
for variables which are null in the true model also hold in the
selected model. We use the classical $t$, $\chi^2$, or $F$
significance tests for single coordinates or groups of coordinates
of $\beta_{A_m}$, and compute $p$-values from the null distributions
truncated to the model selection region. For a variable $j \in A_m$,
the selective null hypothesis and type 1 error are
\begin{equation}
  \label{eq:nullhyp}
  \begin{aligned}
  & H_0(A_m,j) : \beta_{A_m,j} = 0 \\
  & \mathbb P_{m, H_0(A_m,j)}(\text{reject } H_0(A_m,j)|M(y) = m)
  \end{aligned}
\end{equation}
where the probability under $m$ and $H_0(A_m,j)$ is computed from
the probability model \eqref{eq:normalmodel} truncated to the region
implied by $M(y) = m$ and with the constraint on $\mu$ implied by
$H_0(A_m, j)$.
Further details on $\chi^2$ and $F$
tests are provided in \cite{loftus2015groups}, and \cite{lee2013exact}
discuss $z$-tests in the setting where model selection is affine
rather than quadratic.
\cite{fithian2014optimal} discuss the advantages of
selective inference and optimality theory in exponential families.

\section{Test error estimates as quadratic forms}
This section shows how we can write test error estimates as quadratic
forms. Thus, when a model is chosen by minimizing these estimates we
can apply the strategy in \cite{loftus2015groups} to conduct inference
conditional on the chosen model. For simplicity we first illustrate
the approach with forward stepwise, writing $s$ to index steps rather
than $\lambda$, and then discuss how the Lasso and other methods fall
within the same framework.

\subsection{A single training-test split}
Before analyzing full cross-validation, we first consider splitting
the data into independent partitions $(y^\tr, X^\tr)$ for training and
$(y^\te, X^\te)$ for estimating test error. For concreteness assume we
run forward stepwise for a fixed number of steps $S$ and observe
\begin{equation}
  \label{eq:trmodel}
  M^\tr(y^\tr) = m \iff y^\tr \in \bigcap_{j \in J} E^\tr_{m,j}
\end{equation}
where $E^\tr_{m,j}$ is an event of the form $\{ z \in \mathbb{R}^{|\tr|} :
z^TQ^\tr_{m,j}z + A^\tr_{m,j} z + b^\tr_{m,j} \geq 0
\}$ and $|\tr|$ is the size of the training set.
For each $s = 1, \ldots, S$ we have a model in the forward stepwise
path with an associated parameter vector $\hat \beta_{m,s}$ given by 
\begin{equation}
  \label{eq:trbeta}
  \hat \beta_{m,s} = (X^\tr_{m,s})^\dagger y^\tr
\end{equation}
Picking the model in this forward stepwise path with the smallest test
error means finding 
\begin{equation}
  \label{eq:terss}
  \hat s := \argmin_s \text{RSS}^\te_{m,s}(y^\te) = \argmin_s \| y^\te - X^\te_{m,s} \hat \beta_{m,s} \|_2^2
\end{equation}
Denote $P_{m,s} := X^\te_{m,s} (X^\tr_{m,s})^\dagger$, so the RSS
criterion above is $\| y^\te - P_{m,s} y^\tr \|_2^2$. The following
Lemma follows from simple algebra and the definitions.
\begin{lem}
  For all $r \neq s$ with $1 \leq r \leq S$, define
  \begin{equation}
    \label{eq:mintestsets}
    E^\te_{m,r} := \{ y :
    \begin{bmatrix}
      y^\tr \\ y^\te 
    \end{bmatrix}^T
    \begin{bmatrix}
      P_{m,s}^T P_{m,s} - P_{m,r}^T P_{m,r} & -(P_{m,s}-P_{m,r})^T \\
      -(P_{m,s}-P_{m,r}) & 0 \\
    \end{bmatrix}
    \begin{bmatrix}
      y^\tr \\ y^\te 
    \end{bmatrix}
    \leq 0 \}
  \end{equation}
  Then conditional on the models fitted on the training set,
  \begin{equation}
    \label{eq:mintestevent}
    \hat s = s \iff y \in \bigcap_{\substack{r=1 \\ r \neq s}}^S E^\te_{m,r}.
  \end{equation}
\end{lem}

Next we combine all the observed inequalities. We can pad the matrices
defining each $E^\tr_{m,j}$ by adding a block of zeroes for $y^\te$,
so that 
\[
y^\tr \in E^\tr_{m,j} \iff y \in E_{m,j}
\]
We have established
\begin{prop}
\label{prop:traintest}
  The model selection event determined by a single training and test
  split decomposes as the following intersection of quadratic
  inequalities
  \begin{equation}
    \label{eq:mintestevent}
     M^\tr(y^\tr)= m \text{ and } \hat s = s 
    \iff
    y \in \bigcap_{j \in J} E_{m,j} \cap \bigcap_{r=1}^S E^\te_{m,j}
  \end{equation}
  where $J$ is the (finite) index set given in \eqref{eq:trmodel}.
\end{prop}

\subsection{$K$-fold cross-validation}
\label{sec:kfoldcv}
To extend the argument to cross-validation we first need more
notation. Let $1 \leq f \leq K$ index CV folds. Write $X^f$ for the
test set and $X^{-f}$ for the training set for fold $f$. The
model $m_f$ is trained on $(y^{-f}, X^{-f})$, and there is an
associated event 
\begin{equation}
  \label{eq:cvmodel}
  M^{f}(y^{-f}) = m_f \iff y^{-f} \in \bigcap_{j \in J_f} E^f_{m,j}.
\end{equation}
For each sparsity $s$
there is a corresponding fit $\hat \beta_{m_f,s}$. Suppose we use
least squares on the active set, so $\hat \beta_{m_f,s} =
(X^{-f}_{m_f,s})^\dagger y^{-f}$. 
We choose $s$ minimizing the cv-RSS, 
\begin{equation}
  \label{eq:cvRSS}
  s = \argmin_s \sum_{f=1}^K \| y^{f} - X^{f}_{m_f,s} \hat \beta_{m_f,s} \|_2^2
\end{equation}
Define
\begin{equation}
  \label{eq:cvproj}
  P_{f,s} := X^{f}_{m_f,s} (X^{-f}_{m_f,s})^\dagger
\end{equation}
The objective in ~\eqref{eq:cvRSS} can be written
\begin{equation}
  \label{eq:cvRSSexpanded}
  \begin{aligned}
    \text{RSS}(s) &:= \sum_{f=1}^K \| y^f - P_{f,s} y^{-f} \|_2^2 \\
    &= \sum_{f=1}^K \left( \| y^f \|_2^2 - (y^f)^TP_{f,s}y^{-f} - (y^{-f})^T P_{f,s}^T y^f + \| P_{f,s} y^{-f} \|_2^2 \right) \\
    &= \| y \|_2^2 - \underbrace{\sum_{f=1}^K (y^f)^TP_{f,s}y^{-f}}_\text{(I)} - \underbrace{\sum_{f=1}^K (y^{-f})^T P_{f,s}^T y^f}_\text{(II)} + \underbrace{\sum_{f=1}^K (y^{-f})^T[(P_{f,s})^TP_{f,s}]y^{-f}}_\text{(III)} \\
   \end{aligned}
\end{equation}
Let $(\cdot)_f$ denote the block matrix with columns corresponding to
the indices of $f$, so 
\begin{equation}
  \label{eq:cvblocks1}
  (P_{f,s})^TP_{f,s} =
  \begin{bmatrix}
    (P_{f,s})^T_1(P_{f,s})_1 & \cdots & (P_{f,s})^T_1(P_{f,s})_K \\
    \vdots & \ddots & \\
    (P_{f,s})^T_K(P_{f,s})_1 & \cdots & (P_{f,s})^T_K(P_{f,s})_K
  \end{bmatrix}
\end{equation}
\begin{equation}
  \label{eq:cvblocks2}
  P_{f,s}y^{-f} = \sum_{g \neq f} (P_{f,s})_g y^g, \quad
  (y^{-f})^T (P_{f,s})^T = \sum_{g \neq f} (y^g)^T (P_{f,s})_g^T.
\end{equation}

We have found~\eqref{eq:cvRSSexpanded} is a quadratic form which we
can write blockwise with the blocks given by folds. Dropping the $\| y
\|_2^2$ term, we see that for block $p$ the diagonal terms $(y^p)^T
Q_{pp} y^p$ appear precisely in terms from (III) of the form $(y^p)^T
(P_{h,s})^T_p(P_{h,s})_p y^p$ for $h \neq p$. So 
\begin{equation}
  \label{eq:cvquaddiag}
  Q^s_{ff} := \sum_{g\neq f}(P_{g,s})_f^T(P_{g,s})_f, 
\end{equation}
For $q \neq p$, the $pq$ terms in (III) appear as
$(y^p)^T(P_{h,s})^T_p(P_{h,s})_q y^q$ for all $h \notin \{ p, q
\}$. The only $pq$ term in (I) occurs when $f = p$ and equals $(y^p)^T
(P_{p,s})_q y^q$. Similarly, the only $pq$ term in (II) occurs when $f
= q$ and equals $(y^p)^T (P_{q,s})^T_p y^q$. Hence for $f \neq g$ we
define 
\begin{equation}
  \label{eq:cvquadoffd}
  Q^s_{fg} := -(P_{f,s})_g - (P_{g,s})^T_f + \sum_{\substack{h=1\\h \notin \{f, g \}}}^K (P_{h,s})^T_f (P_{h,s})^T_g
\end{equation}

 Let $y_K$ denote observations reordered according to the CV folds. We
 have shown that 
 \begin{equation}
   \label{eq:RSSquad}
   \text{RSS}(s) = \|y\|_2^2 + y_K^T Q^s y_K.
 \end{equation}
This allows us to characterize the cross-validation selection event.
\begin{lem}
Conditional on the models $m_f$ fitted on each training set,
the intersection of quadratic events
\begin{equation}
  \label{eq:cvquad}
  \hat s = s \iff
  \bigcap_{\substack{r=1\\r\neq s}}^S \{ y : y_K^T(Q^s-Q^r)y_K \leq 0 \}
\end{equation}
\end{lem}
Since reordering the observations is accomplished by a permutation
matrix $y_K = Py$, we conclude that the cross-validation selection
procedure is characterized by quadratic inequalities. 
\begin{prop}
\label{prop:cv}
  Define $E_j$ by including the events in \eqref{eq:cvmodel} into
  $\mathbb R^n$, and let $E^\cv_r := \{ z : z^TP^T(Q^r-Q^s)Pz \geq 0
  \}$, then
  \begin{equation}
    \label{eq:mincvevent}
    M^f(y^f) = m_f \text{ for } f = 1, \ldots, K \text{ and } \hat s = s \iff
    y \in \bigcap_{j \in J} E_j \cap \bigcap_{r=1}^S E^\cv_r
  \end{equation}
\end{prop}

Propositions~\ref{prop:traintest} and \ref{prop:cv} allow us to adjust
significance tests for variables included in the final model with
sparsity level $\hat s$. Since the model selection events decompose as
intersections of quadratic inequalities, given a distributional
assumption on $y$ we can condition to the selection event by simple
operations of intersection and solving quadratics. We next discuss how
the Lasso and other methods fit in this framework, and demonstrate some
specific examples of inference conditional on this kind of model
selection.

\subsection{Extensions and applications}

\begin{itemize}
\item Simplifying assumptions such as $K$ dividing $n$ and all folds
  having equal size were for notational ease only, and are not assumed
  in our software implementation.
\item The model selection events of Lasso, forward stepwise, forward
  stepwise with groups, marginal screening, and many other methods fit
  in the quadratic framework \eqref{eq:quaddecomp}.
  Any such method can be used to determine
  the events in \eqref{eq:trmodel} or \eqref{eq:cvmodel}.
\item Similarly, the predictions of various methods, including the
  Lasso, forward stepwise, and others, result in
  various forms of the ``hat matrices'' \eqref{eq:cvproj}. For the
  Lasso there are additional constant terms appearing in the fitted
  values which must be added into the RSS criterion. But these are
  constant on the model selection event, so the approach here still
  works and the only change is that the inequality defining
  $E^\cv_\lambda$ has additional constants.
\item For the Lasso a grid of $\lambda$ values
  can be used instead of the steps $s$ of forward stepwise. Small
  modifications allow stagewise fitting such as \texttt{lar}
  \citep{efron2004lar} or \texttt{glmnet} \citep{glmnet}.
\item The RSS criteria can be penalized in various
  ways to account for differences in model size if desired. For
  example, let $\lambda_{m,s} = 2|m_s|\log(n/K)$ be the BIC penalty
  with $|m_s|$ denoting the number of nonzero parameters in model $m$
  at step $s$ and $n/K$ the number of observations in the test
  set. Then we only need to add the constants $\lambda_{m,s} -
  \lambda_{m,r}$ to the left hand side of the corresponding quadratic
  inequalities. 
\item When $\sigma^2$ is unknown, there is a choice between using
  selective $t$ or $F$ tests or plugging in an estimate of $\sigma$.
  In the latter case, there are estimates that can be computed with
  cross-validation such as those discussed in \cite{reid2013sigma}.
\end{itemize}

\section{Simulations}

\begin{figure}[h!]
  \centering
  \includegraphics[width=\textwidth]{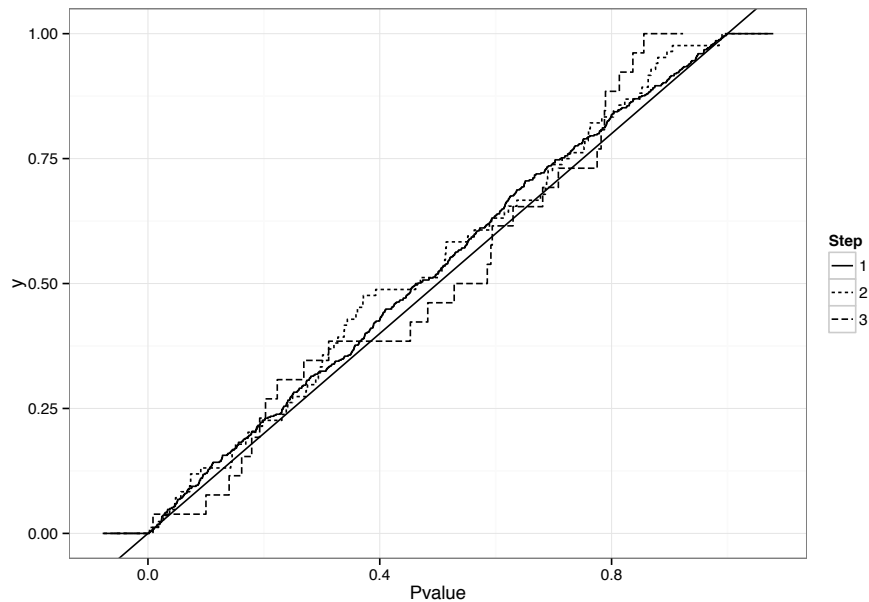}
  \caption{\footnotesize Empirical CDF of selective $p$-values in a
    global null simulation
  with $n = 50$ observations of $p = 100$ independent Gaussian predictors.
  Cross-validation rarely chose models with sparsity larger than three,
  so only the first three steps are plotted.}
  \label{fig:globalnull}
\end{figure}

\begin{figure}[h!]
  \centering
  \includegraphics[width=\textwidth]{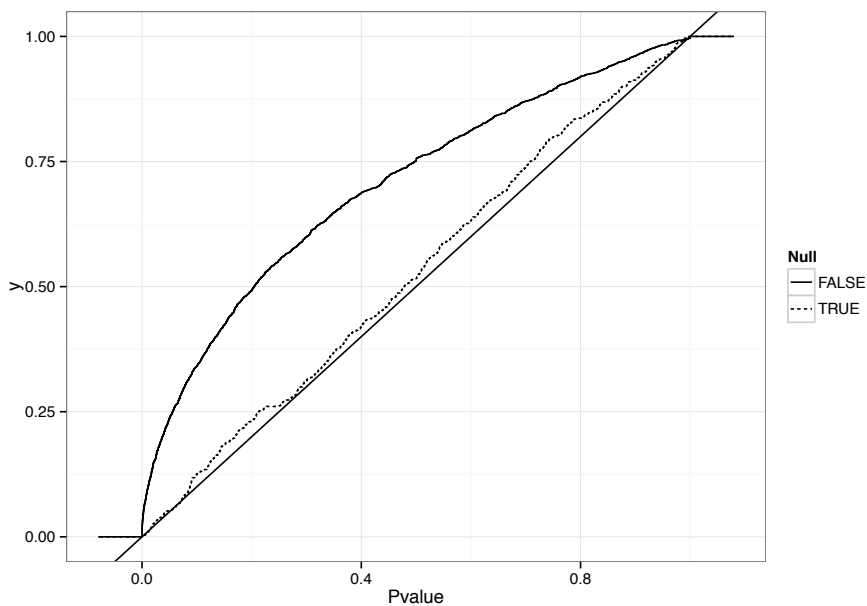}
  \caption{\footnotesize Empirical CDF of selective $p$-values in a
  simulation similar to Fig.~\ref{fig:globalnull} but with true
  sparsity equal to five. The solid line shows that $p$-values
  corresponding to truly nonzero coefficients are small, so the test
  has power. The nonzero coefficients were all equal to $\pm 1$.}
  \label{fig:fivesparse}
\end{figure}

To demonstrate the applicability and power of this method we conducted
simulations using forward stepwise as the model selection procedure as
described in Section~\ref{sec:kfoldcv}. Figures~\ref{fig:globalnull}
and~\ref{fig:fivesparse} show the empirical CDFs of selective $p$-values
computed from truncated $\chi$ tests for the variables included in final
models with sparsity level determined by cross-validation. These figures
show that significance tests for the variables which are null in the true
model have the desired type 1 error control, and significance tests for
the variables which are nonnull in the true model have reasonable power.

\section{Discussion}

The main drawback of our method is that it is computationally expensive.
This cost is mostly due to the complicated geometry of the quadratic
model selection regions \eqref{eq:quaddecomp}. Important special cases,
such as forward stepwise and the Lasso without any groups of variables,
reduce to simpler polyhedral selection regions and this can be exploited
by specialized implementations. This was not explored in the present
work but will be included in a future version of the
\texttt{selectiveInference} R package \cite{SIR}.

There are other limitations associated with the selective inference
approach, but these are not particular to the present work on
cross-validation. Perhaps the greatest of these limitations is that
selective hypotheses may not be in correspondence with hypotheses
about the true model when the model selection procedure performs
poorly. By allowing the use of cross-validation---which is
empirically known to perform quite well---in selective inference,
the present work reduces the severity of this limitation.

Finally, the author is not aware of any previous work analyzing
cross-validation through the quadratic form structure
\eqref{eq:cvquaddiag}--\eqref{eq:RSSquad}. Using this structure to
obtain other new results on cross-validation unrelated to selective
inference---for example, developing theory about the bias of minimum
cross-validation error---is an area of ongoing work. 

\ 

\textbf{Acknowledgement:} The author would like to thank Jonathan Taylor and Robert Tibshirani for helpful comments and suggestions.



\bibliographystyle{imsart-nameyear.bst}
\bibliography{biblio}

\begin{thebibliography}{12}

\bibitem[\protect\citeauthoryear{{Belloni}, {Chernozhukov} and
  {Wang}}{2010}]{sqrtlasso}
\begin{barticle}[author]
\bauthor{\bsnm{{Belloni}},~\bfnm{A.}\binits{A.}},
  \bauthor{\bsnm{{Chernozhukov}},~\bfnm{V.}\binits{V.}} \AND
  \bauthor{\bsnm{{Wang}},~\bfnm{L.}\binits{L.}}
(\byear{2010}).
\btitle{{Square-Root Lasso: Pivotal Recovery of Sparse Signals via Conic
  Programming}}.
\bjournal{ArXiv e-prints}.
\end{barticle}
\endbibitem

\bibitem[\protect\citeauthoryear{Efron et~al.}{2004}]{efron2004lar}
\begin{barticle}[author]
\bauthor{\bsnm{Efron},~\bfnm{Bradley}\binits{B.}},
  \bauthor{\bsnm{Hastie},~\bfnm{Trevor}\binits{T.}},
  \bauthor{\bsnm{Johnstone},~\bfnm{Iain}\binits{I.}},
  \bauthor{\bsnm{Tibshirani},~\bfnm{Robert}\binits{R.}} \betal{et~al.}
(\byear{2004}).
\btitle{Least angle regression}.
\bjournal{The Annals of statistics}
\bvolume{32}
\bpages{407--499}.
\end{barticle}
\endbibitem

\bibitem[\protect\citeauthoryear{Fithian, Sun and
  Taylor}{2014}]{fithian2014optimal}
\begin{barticle}[author]
\bauthor{\bsnm{Fithian},~\bfnm{William}\binits{W.}},
  \bauthor{\bsnm{Sun},~\bfnm{Dennis}\binits{D.}} \AND
  \bauthor{\bsnm{Taylor},~\bfnm{Jonathan}\binits{J.}}
(\byear{2014}).
\btitle{Optimal inference after model selection}.
\bjournal{arXiv preprint arXiv:1410.2597}.
\end{barticle}
\endbibitem

\bibitem[\protect\citeauthoryear{Friedman, Hastie and
  Tibshirani}{2010}]{glmnet}
\begin{barticle}[author]
\bauthor{\bsnm{Friedman},~\bfnm{Jerome}\binits{J.}},
  \bauthor{\bsnm{Hastie},~\bfnm{Trevor}\binits{T.}} \AND
  \bauthor{\bsnm{Tibshirani},~\bfnm{Robert}\binits{R.}}
(\byear{2010}).
\btitle{Regularization Paths for Generalized Linear Models via Coordinate
  Descent}.
\bjournal{Journal of Statistical Software}
\bvolume{33}
\bpages{1--22}.
\end{barticle}
\endbibitem

\bibitem[\protect\citeauthoryear{{Gross}, {Taylor} and
  {Tibshirani}}{2015}]{gross2015internal}
\begin{barticle}[author]
\bauthor{\bsnm{{Gross}},~\bfnm{S.~M.}\binits{S.~M.}},
  \bauthor{\bsnm{{Taylor}},~\bfnm{J.}\binits{J.}} \AND
  \bauthor{\bsnm{{Tibshirani}},~\bfnm{R.}\binits{R.}}
(\byear{2015}).
\btitle{{A Selective Approach to Internal Inference}}.
\bjournal{ArXiv e-prints}.
\end{barticle}
\endbibitem

\bibitem[\protect\citeauthoryear{Lee et~al.}{2015}]{lee2013exact}
\begin{barticle}[author]
\bauthor{\bsnm{Lee},~\bfnm{Jason~D}\binits{J.~D.}},
  \bauthor{\bsnm{Sun},~\bfnm{Dennis~L}\binits{D.~L.}},
  \bauthor{\bsnm{Sun},~\bfnm{Yuekai}\binits{Y.}} \AND
  \bauthor{\bsnm{Taylor},~\bfnm{Jonathan~E}\binits{J.~E.}}
(\byear{2015}).
\btitle{Exact post-selection inference with the lasso}.
\bjournal{Ann. Statist.}
\bnote{To appear}.
\end{barticle}
\endbibitem

\bibitem[\protect\citeauthoryear{{Loftus} and
  {Taylor}}{2015}]{loftus2015groups}
\begin{barticle}[author]
\bauthor{\bsnm{{Loftus}},~\bfnm{J.~R.}\binits{J.~R.}} \AND
  \bauthor{\bsnm{{Taylor}},~\bfnm{J.~E.}\binits{J.~E.}}
(\byear{2015}).
\btitle{{Selective inference in regression models with groups of variables}}.
\bjournal{ArXiv e-prints}.
\end{barticle}
\endbibitem

\bibitem[\protect\citeauthoryear{Reid, Tibshirani and
  Friedman}{2013}]{reid2013sigma}
\begin{barticle}[author]
\bauthor{\bsnm{Reid},~\bfnm{Stephen}\binits{S.}},
  \bauthor{\bsnm{Tibshirani},~\bfnm{Robert}\binits{R.}} \AND
  \bauthor{\bsnm{Friedman},~\bfnm{Jerome}\binits{J.}}
(\byear{2013}).
\btitle{A study of error variance estimation in lasso regression}.
\bjournal{arXiv preprint arXiv:1311.5274}.
\end{barticle}
\endbibitem

\bibitem[\protect\citeauthoryear{Taylor, Loftus and
  Tibshirani}{2015}]{taylor2013tests}
\begin{barticle}[author]
\bauthor{\bsnm{Taylor},~\bfnm{Jonathan~E.}\binits{J.~E.}},
  \bauthor{\bsnm{Loftus},~\bfnm{Joshua~R.}\binits{J.~R.}} \AND
  \bauthor{\bsnm{Tibshirani},~\bfnm{Ryan~J.}\binits{R.~J.}}
(\byear{2015}).
\btitle{Tests in adaptive regression via the Kac-Rice formula}.
\bjournal{Ann. Statist.}
\bnote{To appear}.
\end{barticle}
\endbibitem

\bibitem[\protect\citeauthoryear{Tian, Loftus and
  Taylor}{2015}]{tian2015sqrtlasso}
\begin{barticle}[author]
\bauthor{\bsnm{Tian},~\bfnm{Xiaoying}\binits{X.}},
  \bauthor{\bsnm{Loftus},~\bfnm{Joshua~R}\binits{J.~R.}} \AND
  \bauthor{\bsnm{Taylor},~\bfnm{Jonathan~E}\binits{J.~E.}}
(\byear{2015}).
\btitle{Selective inference with unknown variance via the square-root LASSO}.
\bjournal{arXiv preprint arXiv:1504.08031}.
\end{barticle}
\endbibitem

\bibitem[\protect\citeauthoryear{{Tibshirani}
  et~al.}{2014}]{tibshirani2014exact}
\begin{barticle}[author]
\bauthor{\bsnm{{Tibshirani}},~\bfnm{R.~J.}\binits{R.~J.}},
  \bauthor{\bsnm{{Taylor}},~\bfnm{J.}\binits{J.}},
  \bauthor{\bsnm{{Lockhart}},~\bfnm{R.}\binits{R.}} \AND
  \bauthor{\bsnm{{Tibshirani}},~\bfnm{R.}\binits{R.}}
(\byear{2014}).
\btitle{{Exact Post-Selection Inference for Sequential Regression Procedures}}.
\bjournal{ArXiv e-prints}.
\end{barticle}
\endbibitem

\bibitem[\protect\citeauthoryear{Tibshirani et~al.}{2015}]{SIR}
\begin{bmanual}[author]
\bauthor{\bsnm{Tibshirani},~\bfnm{Ryan}\binits{R.}},
  \bauthor{\bsnm{Tibshirani},~\bfnm{Rob}\binits{R.}},
  \bauthor{\bsnm{Taylor},~\bfnm{Jonathan}\binits{J.}},
  \bauthor{\bsnm{Loftus},~\bfnm{Joshua}\binits{J.}} \AND
  \bauthor{\bsnm{Reid},~\bfnm{Stephen}\binits{S.}}
(\byear{2015}).
\btitle{selectiveInference: Tools for Selective Inference}
\bnote{R package version 1.1.1}.
\end{bmanual}
\endbibitem

\end{thebibliography}

\end{document}